\newcommand{\beq}{\begin{equation}}
\newcommand{\eeq}{\end{equation}}
\newcommand{\beqa}{\begin{eqnarray}}
\newcommand{\eeqa}{\end{eqnarray}}
\newcommand{\ket}[1]{| #1 \rangle}
\newcommand{\bra}[1]{\langle #1 |}

\newcommand{\opa}{\hat{a}}
\newcommand{\ma}{a_0}
\newcommand{\opb}{\hat{b}}
\newcommand{\mb}{b_0}
\newcommand{\opc}{\hat{c}}
\newcommand{\mc}{c_0}
\newcommand{\opd}{\hat{d}}
\newcommand{\md}{d_0}
\newcommand{\ope}{\hat{e}}
\newcommand{\opf}{\hat{f}}
\newcommand{\mf}{f_0}
\newcommand{\opg}{\hat{g}}
\newcommand{\oph}{\hat{h}}
\newcommand{\opk}{\hat{k}}
\newcommand{\mk}{k_0}
\newcommand{\opr}{\hat{r}}
\newcommand{\oprp}{\hat{r}^{'}}
\newcommand{\opu}{\hat{u}}
\newcommand{\opv}{\hat{v}}
\newcommand{\opap}{\hat{a}^{'}}
\newcommand{\map}{a_0^{'}}

\newcommand{\opfp}{\hat{f}^{'}}
\newcommand{\mfp}{f_0^{'}}
\newcommand{\opkp}{\hat{k}^{'}}
\newcommand{\opvp}{\hat{v}^{'}}

\newcommand{\oprho}{\hat{\rho}}

\newcommand{\mr}{r_0}
\newcommand{\mrp}{r_0^{'}}
\newcommand{\aone}{\hat{a}_1}
\newcommand{\atwo}{\hat{a}_2}
\newcommand{\rone}{\hat{r}_1}
\newcommand{\rtwo}{\hat{r}_2}
\newcommand{\infig}[2]{
\vskip 5mm
\begin{center}
  \resizebox{#2}{!}{\includegraphics*{#1}}
\end{center}
}

\documentclass[pra,twocolumn,showpacs]{revtex4}
\usepackage{graphicx}

\begin{document}


\title{Mutual first order coherence of phase-locked lasers}

\author{Hoshang Heydari}
\email{hoshang@ele.kth.se} \homepage{http://www.ele.kth.se/QEO/}
\affiliation{Department of Microelectronics and Information
Technology, Royal Institute of Technology (KTH), Electrum 229,
SE-164 40 Kista, Sweden}

\author{Gunnar Bj\"{o}rk}
\affiliation{Department of Microelectronics and Information
Technology, Royal Institute of Technology (KTH), Electrum 229,
SE-164 40 Kista, Sweden}

\date{\today}

\begin{abstract}
We argue that (first-order) coherence is a relative, and not an
absolute, property. It is shown how feedforward or feedback can be
employed to make two (or more) lasers relatively coherent. We also
show that after the relative coherence is established, the two
lasers will stay relatively coherent for some time even if the
feedforward or feedback loop has been turned off, enabling, e.g.,
demonstration of unconditional quantum teleportation using lasers.
\end{abstract}


\maketitle

\section{Introduction}
Recently, several authors have pointed out that although it has
been known for a long time that a laser does not generate a
coherent state, it is still possible to use laser light to observe
phenomena that are theoretically described in terms of optical
(first order) coherence \cite{Molmer,Rudolph,Enk,Wiseman}.

M\o lmer \cite{Molmer} argues that optical coherence, manifested,
e.g., in a non-vanishing expectation value of the electric field
operator, is a ``convenient fiction'' and he demonstrates that
experiments that are usually interpreted in terms of optical
coherence need not be described in those terms. Moreover, he
argues that optical coherence is not easily generated, and in
particular, that a laser does not generate coherent state. Rudolph
and Sanders \cite{Rudolph} show that if M\o lmer's assertions are
true, so that the output state of a laser is given by the density
operator \beq \hat{\rho}_L=\int_0^{2 \pi} {d \phi \over 2 \pi}
\ket{\alpha e^{i \phi}}\bra{\alpha e^{i \phi}} = e^{- \alpha^2}
\sum_{n=0}^\infty {\alpha^{2 n}\over n !} \ket{n} \bra{n},
\label{eq: mixed state} \eeq then quantum teleportation is not
possible under the conditions postulated for ``unconditional
quantum teleportation'' \cite{Furusawa,Braunstein}. Rudolph and
Sanders do not rule out the existence of sources of coherent
states, but argue that lasers do not produce light with (first
order) optical coherence. Van Enk and Fuchs \cite{Enk} claim that
while a laser does not produce optical coherence in M\o lmer's
sense, subsequent temporal modes of the laser output are phase
correlated. That is, if the output from a laser is expanded in
subsequent orthogonal temporal modes, the state of $M$ such modes
is \beq \int_0^{2 \pi} {d \phi \over 2 \pi} \ket{\alpha e^{i
\phi}}\bra{\alpha e^{i \phi}}^{\otimes M} , \label{eq: enk}\eeq
where the symbol $^{\otimes M}$ denotes a $M$-fold tensor product.
A requirement for this is that the laser's coherence time is
longer than $M$ times the time duration of a temporal mode. In
such a state all the $M$ subsequent modes are first order coherent
relative to each other, in contrast to the $M$-mode state
$\hat{\rho}_L^{\otimes M}$. Both states have a vanishing
expectation value of the electric field operator for all modes.
Van Enk and Fuchs claim that such ``phase coherence'' is
sufficient to allow unconditional quantum teleportation. Wiseman
\cite{Wiseman}, finally, asserts that there are ``no devices that
can generate `true coherence' any better than a laser.'' The basis
for his claim is that any oscillator, or oscillation, can only be
described relative to some accepted ``clock'' standard. He goes on
stating that nothing suggests that there exist \textit{any} better
``clock'' at optical frequencies than a laser. We subscribe to
this view, and it is compounded by the fact that at the National
Institute of Standards i Boulder, Colorado, the next generation of
an ``atomic clock'' in development is indeed an optical clock
\cite{Diddams}. That is, the clock oscillator operates at an
optical-, rather than a microwave-transition of an atom.

The coherent state $\ket{e^{i \phi}|\alpha|}$ of an oscillator
with the angular center frequency $\omega_0$, can mathematically
be generated from the vacuum state through the displacement
operator. It is a minimum uncertainty state in the in-phase
operator $\aone = (\hat{a}+\hat{a}^\dagger)/2$ and the
quadrature-phase operator $\atwo = (\hat{a}-\hat{a}^\dagger)/2i$,
where $\hat{a}$ ($\hat{a}^\dagger$) is the bosonic annihilation
(creation) operator. The quadrature amplitude fluctuation
operators are defined $\triangle \opa_i = \opa_i - \langle
\opa_i\rangle$, with $i = 1, 2$. In Table \ref{table: variances}
we have computed the expectation values of the quadrature
amplitudes, the photon number, and their variances for the two
states $\hat{\rho}_L$ and $\ket{e^{i \phi}|\alpha|}$ (for the
particular choice $\phi=0$) in columns two and three. The table
clearly shows that the observable statistics of the two states
differ significantly. Hence, they cannot both describe the output
of a laser, at least not if similar initial and boundary
conditions for the laser and the detector are assumed. Wang {\it
et al.} have also studied and quantified this difference
\cite{Wang}.

In the following we shall describe a simple model of phase locking
of two independent lasers to a master laser. Phase-locking of two
lasers has already been demonstrated using both CW \cite{Ye} and
pulsed lasers \cite{Shelton}. However, phase-locking of two
independent lasers close to the quantum limit, that is such that
the relative phase fluctuations are comparable to those of two
ideal coherent states, has never been demonstrated. We shall see,
that it is possible to stabilize two lasers so that they are
relatively first-order coherent to that extent, although they are
incoherent relative to any third, auxiliary laser, or other
``clock standard''. A similar proposal has also been put forth by
Fujii \cite{Fujii}, but in \cite{Fujii}, the analysis is geared
towards a practical implementation of quantum teleportation,
whereas we direct our interest towards the question whether first
order coherence between two lasers can be established at all.

It will be convenient to work in the Heisenberg picture in the
spectral domain. Since a CW laser output consists of a continuous
photon flux, we shall work with the photon flux operator
$\hat{r}$. (See subsection \ref{app: expectation values} in the
Appendix for a detailed discussion of $\hat{r}$.) For a coherent
state, the spectral relations corresponding to equations three and
four in the third column of Table \ref{table: variances} are
\cite{Yamamoto}
\begin{eqnarray}
S_{\Delta \rone} (\Omega) & = & 1/4 , \label{eq: spectrum 1}\\
S_{\Delta \rtwo} (\Omega)  & = & 1/4 , \label{eq: spectrum 2}
\end{eqnarray}
where $\Omega=\omega - \omega_0$ so that, e.g., $S_{\Delta \rone}
(\Omega)$ is the double-sided power spectrum of $\Delta \rone$
around the center frequency $\omega_0$. (Note that the laser
external field spectra computed in Ref. \cite{Yamamoto} are
single-sided spectra.) These relations demonstrate a particular
feature of the coherent state, namely a frequency independent
quadrature amplitude noise spectrum. Operationally, this means
that the quadrature amplitude noise of a coherent state is
stationary. Moreover, a field in a coherent state remains in a
coherent state independent of the detector temporal response
function (see the Appendix).

The corresponding spectra for the laser external field have also been
computed in \cite{Yamamoto}. While the model employed was
specifically targeting a semiconductor laser, the general features
are largely independent of the laser type. It was found that the
corresponding relations for (a somewhat idealized) laser, pumped
high above the threshold, are
\begin{eqnarray}
S_{\Delta \rone} (\Omega) & = & 1/4 ~\text{and} \label{eq: laser spectrum 1}\\
S_{\Delta \rtwo} (\Omega)  & = & {\Omega^2 + \gamma^2\over
4\Omega^2} ,
\label{eq: laser spectrum 2}
\end{eqnarray}
where $\gamma$ is the inverse of the laser photon lifetime. We see
that (\ref{eq: laser spectrum 2}) radically differs from (\ref{eq:
spectrum 2}) in that Eq. (\ref{eq: laser spectrum 2}) has a
$1/\Omega^2$ behavior at low frequencies. This is characteristic
of a Wiener process, leading to diffusion. Therefore, even if the
phase of the external field amplitude was known (e.g., through a
series of measurements) at some time $t_0$, the phase will be
randomized through phase diffusion at times $t \gg t_0 +
1/\gamma$. Therefore, unless the field of the laser has been
measured relative to a reference within the laser's coherence time
$1/\gamma$, the state of a laser is well described by the density
operator $\hat{\rho}_L$ irrespective of how relatively stable the
reference oscillator is. The density matrix $\hat{\rho}_L$ also
describes the state of a free-running laser, that is, a laser
whose phase relative to our reference is unknown. This result is
indeed expected from symmetry arguments, and M\o lmer, in
particular, emphasize that there is no mechanism in a free-running
laser that breaks the in-phase, quadrature-phase symmetry. Hence,
the expectation value of the electric field of an unmonitored,
free-running laser vanishes.

\begin{figure}[htbp]
\infig{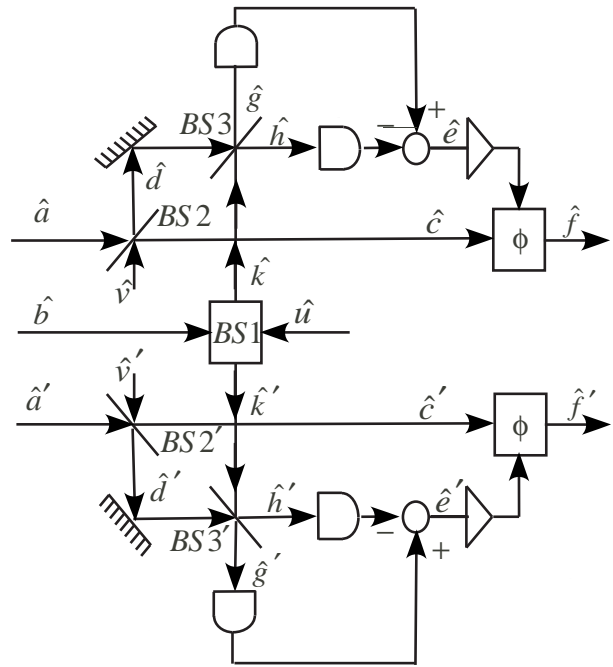}{8cm} \caption{A schematic drawing of two 
lasers locked by feedforward to a master oscillator laser.} 
\label{Fig: Setup}
\end{figure}

\section{Phase-locked lasers}

Now consider the case when we stabilize the laser field with the
help of a master oscillator. In the following we will consider
feedforward stabilization, for simplicity. Later, we shall briefly
discuss feedback stabilization. Fig. \ref{Fig: Setup} shows a
schematic drawing of two lasers locked to a master oscillator by
feedforward. The notation we will use is defined by the figure.
The master oscillator (laser) field $\opb$ is divided into two by
beam splitter one (BS1) which is a 50/50 beam splitter. The field
into the other port of the beam splitter is a vacuum field $\opu$,
so that the beam splitter output modes $\opk$ and $\opkp$ are
given by $\opk=\frac{1}{\sqrt{2}}(\opb+\opu)$ and
$\opkp=\frac{1}{\sqrt{2}}(\opb-\opu)$.

Next, we look at beam splitter two (BS2), where part of the laser
field $\opa$ is tapped to get a locking signal $\opd$. Again, a
vacuum field, denoted $\opv$, is incident on the other input port.
Hence, the two output fields $\opd$ and $\opc$ are given by
\begin{equation}
\opd = r \opa + t \opv, ~\text{and} ~\opc = t \opa - r \opv ,
\end{equation}
where $r=\sin\theta$ ($t=\cos\theta$) is the reflectivity
(transmissivity) of the beam splitter. At beam splitter number
three (BS3), which is also a 50/50 splitter, the fields $\opd$ and
$\opk$ are mixed, and subsequently the outputs $\opg$ and $\oph$,
which are given by
\begin{equation}
    \opg =(\opd + \opk)/\sqrt{2} = (r\opa+t\opv)/\sqrt{2}
    +\frac{1}{2}(\opb+\opu)
\end{equation}
and
\begin{equation}
    \oph=(\opd - \opk)/\sqrt{2}=(r\opa+t\opv)/\sqrt{2}
    -\frac{1}{2}(\opb+\opu)
\end{equation}
are measured by photodetectors. The detector photocurrents will be
proportional to $\opg^\dagger\opg$ and $\oph^\dagger\oph$, that
can be expressed as
\begin{equation}
    \opg^\dagger\opg=( \opd^\dagger\opd + \opd^\dagger\opk
    + \opk^\dagger\opd+ \opk^\dagger\opk )/2
\end{equation}
and
\begin{equation}
    \oph^\dagger\oph= (\opd^\dagger\opd
    -\opd^\dagger\opk
    -\opk^\dagger\opd+\opk^\dagger\opk)/2 .
\end{equation}
The difference between the measured photocurrents
$\opg^\dagger\opg - \oph^\dagger\oph$ will be the homodyne signal
$\ope$, which we call the error signal. This signal is simply
\begin{equation}
   \ope =\opd^\dagger\opk+\opk^\dagger\opd .
\label{eq: Difference} \eeq Now, expand the operators in
quadrature components and assume that $\opd$ leads $\opk$ by the
relative phase angle $\pi/2$. (this choice of relative phase will
beat the excitation of one field with the quadrature amplitude of
the other field.) That is, we multiply the operator $\opd$ with
$\exp(i \pi/2) = i$ and remember that both fields are expressed
relative to the same fiducial field. (Below, we shall see that the
relative phase angle $\pi/2$ is the stable feedforward locking
point.) This will yield the mean error signal zero, and the
fluctuating part of the error signal will be
\begin{eqnarray}
\ope & = & -i(\md+\triangle\opd_{1}-i\triangle\opd_{2})(\mk
   +\triangle\opk_{1}+i\triangle\opk_{2}) \\
 &&  +i(\mk+\triangle\opk_{1}-i\triangle\opk_{2})(\md
   +\triangle\opd_{1}+i\triangle\opd_{2}) \\
 &=& 2(\md\triangle\opk_{2}-\mk\triangle\opd_{2})
   +\mathcal{O}(\triangle^{2}) ,
\label{eq: linearized homodyne}
\end{eqnarray}
where $\mathcal{O}(\triangle^{2})$ denotes terms of second order
in the quadrature flux fluctuation operators.

It is well worth examining Eq. (\ref{eq: linearized homodyne}) in
some detail, since it is the generic noise equation for homodyne
measurements. Assume, that each of the fields $\opk$ and $\opd$ is
in a coherent state. Hence, it follows that the two fields are
relatively coherent. Since the quadrature amplitude fluctuations
of a coherent state are equal, and independent of the state's
excitation, the homodyne detection noise is dominated by the
weaker field, since the weaker field's quadrature amplitude noise
beats against the stronger field's in-phase amplitude. (It is
customary to refer to the stronger field as the ``local
oscillator.'') If the two  fields have unequal quadrature
amplitude fluctuations, e.g., assume that
$S_{\triangle\opk_{2}}(\Omega) > S_{\triangle\opd_{2}}(\Omega)$,
then the homodyne detection noise is still dominated by the
quadrature amplitude fluctuations of $\opd$ provided that $\mk^2 >
\md^2
S_{\triangle\opk_{2}}(\Omega)/S_{\triangle\opd_{2}}(\Omega)$.

Using the relations between incident and output fields above, we
can re-express this equation in the (quadrature expansion of the)
input fields $\opa$, $\opb$, $\opu$, and $\opv$. The result (to
first order in the fluctuation operators) is
\begin{equation}
   \ope =\frac{r
   \ma}{\sqrt{2}}(\triangle\opb_{2}+\triangle\opu_{2})
   -\frac{\mb}{\sqrt{2}}(r\triangle\opa_{2}+t\triangle\opv_{2}) .
\label{eq: error signal}
\end{equation}
This is almost what we desire. Remember that the objective is to
reduce the large fluctuations of $\triangle\opa_{2}$ at low
frequencies. These are the fluctuations leading to phase
diffusion. The error signal clearly contain the information about
these fluctuations. By making the master oscillator flux amplitude
sufficiently large, so that $\mb$ is much larger than $\ma$, this
noise information will dominate the error signal at low
frequencies. (Remember that the spectral density of
$\triangle\opa_{2}$ and $\triangle\opb_{2}$ increases as
$\Omega^{-2}$, whereas the spectral density of $\triangle\opu_{2}$
and $\triangle\opv_{2}$ is 1/4, independent of frequency.) From
the expression (\ref{eq: error signal}) we see that under this
condition, if the relative phase between $\triangle\opa_{2}$ and
$\triangle\opb_{2}$ is larger than $\pi/2$, that is, either
$\triangle\opa_{2}$ is positive or $\triangle\opb_{2}$ is
negative, then the error signal is negative. In order to
compensate for the $\triangle\opa_{2}$ fluctuations, the error
signal should be feed forward, keeping the sign, to a phase
shifter. Let us, therefore, see what happens when we apply some
phase shift, say $\phi$, to an output signal
$\opc=\mc+\triangle\opc_{1}+i\triangle\opc_{2}$. In this case we
denote the phase shifted  signal by $\opf$
\begin{eqnarray}
\opf & = & \exp(i \phi) \opc \\
& = & (\mc + \triangle\opc_{1})\cos\phi - \triangle\opc_{2} \sin\phi \\
&& + i [\triangle\opc_{2} \cos\phi + (\mc + \triangle\opc_{1}) \sin\phi] \\
& \approx & \mc + \triangle\opc_{1} + i (\triangle\opc_{2} + \mc \phi) ,
\end{eqnarray}
where we have linearized the equation to first order in $\phi$,
$\triangle\opc_{1}$, and $\triangle\opc_{2}$. Assume now that the
phase shift $\phi$ is equal to the error signal $\ope$ times some
feedforward gain ${\cal G}$. The output signal after phase
correction by the error signal is given by
\begin{eqnarray}
\opf \approx t(\ma+\triangle\opa_{1}+i\triangle\opa_{2})
-r(\triangle\opv_{1}+i\triangle\opv_{2}) +i t \ma {\cal G} \ope .
\end{eqnarray}
To cancel the noise term $\triangle\hat{a}_{2}$ in the expression
above, we have to adjust the feedforward gain so that \beq {\cal
G} = \sqrt{2}/r \ma \mb . \label{eq: gain} \eeq After doing so,
the output signal becomes
\begin{equation}
\opf = t\ma+t\triangle\opa_{1} - r\triangle\opv_{1} + i \left [
\frac{t \ma}{\mb}(\triangle\opb_{2}+\triangle\opu_{2}) -
\frac{1}{r}\triangle\opv_{2} \right ] , \label{eq: final result}
\end{equation}
where we have used the relation $t^2 + r^2 = 1$. From this
equation we see that the noise term $\triangle\opa_{2}$,
describing the phase-diffusion noise originating in the laser
generating the $\opa$ field, is absent. Instead, a new diffusion
noise term $t \ma \triangle\opb_{2}/\mb$ has appeared. Since the
master oscillator is also a laser, this noise term will also have
a spectral density proportional to $\Omega^{-2}$ at low
frequencies. Hence, one Wiener process, $\triangle\opa_{2}$
(relative to the fiducial reference), has simply been replaced by
another, $\triangle\opb_{2}$.

Let us now see what happens with the output field $\opfp$ if we
subject the laser field $\opap$ to the same sequence of
measurement and feedforward as the field $\opa$. The calculation
of $\opfp$ proceeds in a similar fashion as for $\opf$, except
that the field $\opkp=(\opb - \opu)/\sqrt{2}$. Hence, the only
difference from the expression (\ref{eq: final result}) above is
that all fields should be primed, except for the master oscillator
field $\opb$ and the vacuum field operator $\opu$. However, the
sign in front of the operator $\opu$ should be reversed. Hence,
the output $\opfp$ becomes

\begin{equation}
\opfp = t\map + t\triangle\opap_{1} - r\triangle\opvp_{1} + i
\left [ \frac{t \map}{\mb}(\triangle\opb_{2}-\triangle\opu_{2}) -
\frac{1}{r}\triangle\opvp_{2} \right ] .
\end{equation}
Assume that the two field amplitudes $t\ma$ and $t\map$ are equal
and that the two fields $\opf$ and $\opfp$ are detected by a
balanced homodyne detector. The detector output, that is the
beat-note between the two incident fields $\opf$ and $\opfp$, can
then be deduced in a straightforward fashion by comparing with
(\ref{eq: linearized homodyne}). The detector output fluctuations
will be
\begin{equation}
\mf \triangle\opfp_2 - \mfp \triangle\opf_2 = 2 t \ma
\left(\frac{2 t \ma \triangle\opu_2}{\mb} +
\frac{\triangle\opvp_{2} -\triangle\opv}{r}  \right ). \label{eq:
homodyne noise}
\end{equation}
From this equation we see that the \textit{relative-noise spectrum
is no longer proportional to $\Omega^{-2}$ at low frequencies, but
is flat}, since $\triangle\opu_2$, $\triangle\opv_2$ and
$\triangle\opvp_{2}$ all emanate from vacuum fluctuations and
hence have frequency independent spectra with the spectral density
1/4. Recalling that an earlier assumption was that $\mb \gg \ma$,
the first term on the right hand side of (\ref{eq: homodyne
noise}) can be neglected. Therefore, the spectral density of the
detector fluctuations is, to a good approximation, $2(t \ma/r)^2$
at all frequencies (see the Appendix). In comparison, the homodyne
detector spectra from two coherent states, with field amplitudes
$t \ma$, would be $2(t \ma)^2$. Hence, as $r \rightarrow 1$, the
relative quadrature-phase noise of the two locked lasers will
approach the noise level of two coherent state sources. At the
same time, the mean in-phase amplitude of $\opf$ and $\opfp$ would
approach zero. Assume now that we use the relatively coherent
fields $\opf$ and $\opfp$ as local oscillators in two separate
homodyne measurements. Using (\ref{eq: linearized homodyne}), it
is not difficult to prove that in order to minimize the influence
of the local oscillator quadrature amplitude noise, the choice $t
= r = 2^{-1/2}$ should be made. Assume, for simplicity, that the
field we want to measure is a coherent state with mean amplitude
$\md$. If so, the coherent state's  quadrature amplitude noise
will dominate the measurement fluctuations if $\ma
> 2 \md$. That is, with this choice of mirror transmission $t$,
the two fields $\opf$ and $\opfp$ are relatively coherent, each
with a noise spectral density twice of that of a coherent state.
For a more detailed discussion of the connection between (\ref{eq:
homodyne noise}) and the first order coherence function between
the locked lasers, the reader is referred to the Appendix,
subsection \ref{appendix: first order }.

An additional consequence, discussed in more detail in subsection
\ref{app: laser output} in the Appendix, is that within a time
small compared to the inverse spectral linewidth of the locked
lasers, the lasers will remain relatively coherent even if the
feedforward locking loop is turned off. It will thus be possible
to do, e.g., continuous variable teleportation without giving up
the quantum teleportation criteria given in \cite{Braunstein}

\section{Feedforward v.s. feedback}

Above, we have analyzed the situation where two lasers are phase
locked to a third laser by feedforward. While such a scheme lends
itself to a simple analysis, the scheme has obvious practical
shortcomings. One is the need of a precise control of the
feedforward gain. In order to completely suppress the phase
diffusion of each laser, relative to the master laser, the
feedforward gain must be precisely set according to (\ref{eq:
gain}). This will be impossible in reality. In addition, the mean
absolute value of the error signal (\ref{eq: error signal}) will
be grow with time as $t^{1/2}$, there $t$ is the time since the
main information the error signal contains, $\triangle\opa_{2}$,
is a non-stationary term. The error signal is used to drive a
phase-compensating device, e.g., an electrooptical crystal or an
adjustable piezoelectrically controlled delay line. However, due
to the fact that with time, the error signal will increase without
limit, any phase-compensating device will eventually run out of
range and will have to be reset.

In reality, it would be wiser to lock the lasers by feedback, and
the feedback should act on the laser frequency (e.g., translating
one of the laser's mirrors) after appropriate filtering. In a
feedback loop, the relative fluctuations will initially become
smaller and smaller with increasing feedback gain. If the feedback
measurement and feedback loop is carefully designed, it is
possible to reach the limit set by quantum noise manifested in
(\ref{eq: homodyne noise}) before (non-fundamental) feedback loop
fluctuations are sufficiently amplified to dominate the locked
lasers' output. In addition, since the accumulated relative phase
equals the time integrated frequency difference between the master
and slave laser, the phase-compensating device will not run out of
range. Differently stated, the lasers frequency difference is the
time derivative of the lasers' relative phase. Therefore, if the
latter noise process has a spectrum proportional to $\Omega^{-2}$,
the former noise has a flat spectrum at low frequencies. The
frequency error signal is hence a stationary process. Hence, if
the feedback is implemented by moving one of the laser's mirrors,
its position needs only to be adjusted within a fixed range and
the actuator moving the mirror need not run out of range.

The disadvantage with feedback locking is that it is more
difficult to analyze, and that one risks self oscillation at
baseband frequencies $\Omega$ where $\Omega \tau_f = \pi$. Here,
$\tau_f$ is the feedback loop time delay. At this frequency the
feedback loop does no longer suppress the fluctuation but instead
enhances them. If the feedback loop gain is too high, one induces
self oscillation. Hence, one needs to make a compromise between
the feedback loop stability and the feedback gain, all while
maximizing the feedback loop bandwidth. Since our analysis
essentially only had the purpose to point out that coherence of a
harmonic oscillator must always be relative to some reference
oscillator, we will not delve deeper into these issues here.

\section{Conclusions}

We have argued that first order coherence is a relative property,
and that if one accepts this view, the light emanating from a
laser can be first-order coherent relative to itself at an earlier
time, or to a different laser. Due to phase diffusion, two lasers
will only stay relatively coherent for a time smaller than the
smallest of inverses the lasers' respective linewidth unless they
are actively phase-locked. Since, in principle, the laser
linewidth can be as small as one wishes, this fact does not
prevent lasers to be used to demonstrate, e.g., unconditional
quantum teleportation. It could in principle be done by locking
Alice's and Bob's homodyning lasers to a master laser before the
unknown state and the shared entangled are measured with the help
of the laser. The locking is then turned off, and Alice's and
Bob's respective laser become free running. The two
experimentalists now have roughly the lasers' coherence time to
perform the teleportation, including the Alice's homodyne
measurements, the classical communication, and the final unitary
evolution by Bob. A typical laser's coherence time may seem short
on the human time-scale ($\sim$1 s). However, on an electronics
time-scale ($\sim$1 ns), a two good lasers stay relatively
coherent for a long time.

In the Appendix, we have included a discussion why the popular
model of laser fluctuations using Langevin noise sources fail to
describe the laser's behavior for times long compared to the
laser's inverse linewidth. The linearization customarily employed
in the solution of these equations neglect to take the restoring
force of the quadrature-phase amplitude fluctuations in account,
and hence these equations erroneously predicts that it is only the
quadrature-phase amplitude, relative to some fix fiducial
reference, that diffuses. Unfortunately, if the equations are not
linearized, they are difficult to solve analytically.

\begin{acknowledgments}
We would like to thank Dr. Jonas S\"{o}derholm and Professor
Shuichiro Inoue for valuable discussions. The work was supported
by the Swedish Research Council (VR) and the Swedish Foundation
for Strategic Research (SSF).
\end{acknowledgments}

\appendix

\section{Expectation values from flux spectra}

\subsection{Expectation values of a coherent state}
\label{app: expectation values}

Here we derive the relation between the external field operator
$\hat{r}$ and the mean photon number and the photon number
variance of a specific mode. The external field operator $\hat{r}$
correspond to a photon flux amplitude. The photon flux operator is
hence given by $\hat{r}^\dagger \hat{r}$, and has the unit Hz. The
expectation value of $\hat{r}^\dagger \hat{r}$ gives the mean
photon flux, in photons per second. Let us expand $\hat{r}$
relative some fiducial reference as $\hat{r}(t)=\mr + \Delta
\rone(t) + i \Delta \rtwo(t)$. Here, the field amplitude $\mr$ has
been assumed to be in phase with a fiducial reference. Expanding
to first order in the fluctuation operators $\Delta \rone(t)$ and
$\Delta \rtwo(t)$ we get \beq \hat{r}^\dagger \hat{r} \approx
\mr^2 + 2 \mr \Delta \rone(t) . \eeq Assume now that a
photodetector temporal mode-function $h$ is a rectangle function
of duration $T$, see Fig. \ref{Fig: Mode function}. The detector
output, as a function of time, will be the convolution between the
flux and the detector response function \beq R(t)=(\hat{r}^\dagger
\hat{r}) \ast h \approx [\mr^2 + 2 r_0 \Delta \rone(t)] \ast h .
\eeq

\begin{figure}[htbp]
\infig{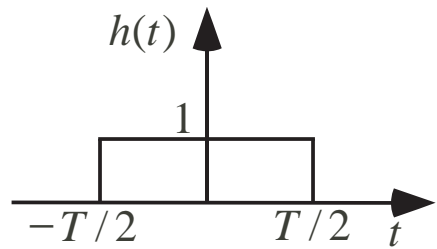}{6cm} \caption{The assumed temporal mode 
function.} \label{Fig: Mode function}
\end{figure}
Sampling this function every $T$ seconds, yields the photon 
number in subsequent orthogonal temporal modes. The detected mean 
photon number per mode hence becomes \beq \langle \hat{n} \rangle 
= r_0^2 T . \label{eq: mean photon number} \eeq We would now like 
to compute the photon number variance of such a mode. Since the 
coherent field flux fluctuations are ergodic we can proceed as 
follows: Define $y(t)=\Delta \rone(t) \ast h$. The spectral 
density of this signal is \beq S_{y(t)}(\Omega)=S_{\rone(t)} 
|H(\Omega)|^2 = {1 \over 4} |H(\Omega)|^2 , \eeq where the 
spectral filter function $H(\Omega)$ is the Fourier transform of 
$h$ and we have used the fact that the spectral density of the 
quadrature amplitude fluctuations of a coherent state flux is 1/4 
according to (\ref{eq: spectrum 1}). The first order correlation 
function $G(\tau)$ of $y(t)$ is the (inverse) Fourier transform 
of its power spectrum, so that \beq G_{y(t)}(\tau) = {\cal 
F}^{-1} S_{y(t)}(\Omega) = {1 \over 4} h \ast h , \eeq where we 
have used the fact that the Fourier transform of a product of 
functions is the convolution of their respective inverses. 
Finally, the variance of $y(t)$ is $G_{y(t)}(0)$. Looking at 
figure \ref{Fig: Mode function convolution}, we find that 
$G_{y(t)}(0)= T/4$. This means that the photon number variance 
can be expressed \beq \langle \Delta \hat{n}^2 \rangle = 4 \mr^2 
\langle (\Delta \rone(t)) \ast h)^2 \rangle = 4 \mr^2 T/4 = T 
\mr^2 . \eeq Comparing with (\ref{eq: mean photon number}) we see 
that we arrive at the expected result for a coherent state.
\begin{figure}[htbp]
\infig{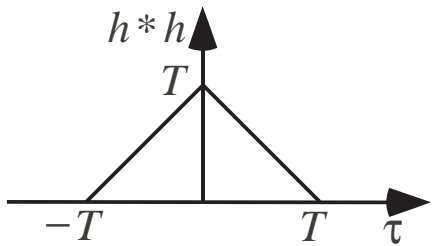}{6cm}\caption{The temporal mode function 
convolution.} \label{Fig: Mode function convolution}
\end{figure}

To derive the mean quadrature amplitudes and their variances we
proceed in a similar way. Suppose two fields, characterized by the
photon flux amplitudes $\opr$ and $\oprp$ are incident on a
homodyne detector. If the two fields are relatively coherent and
have the same phase, we can use (\ref{eq: Difference}) and insert
the expansion of the fields in quadrature amplitude components
(e.g., $\opr = \mr + \triangle \opr_1 + i \triangle \opr_2$). To
the first order in the fluctuation operators the detector output
becomes: \beq 2 \mr \mrp + \mr \triangle \oprp_1 + \mrp \triangle
\opr_1 + O(\triangle^2). \label{eq: hom signal} \eeq Passing this
signal trough appropriate detector filter, (integrate for $T$
seconds and then dump), the mean detection signal becomes \beq 2
\mr \mrp T . \eeq The prefactor 2 in the equation above is simply
a scaling factor. Since the signal emanates from the projection of
one in-phase amplitude on another, we conclude from the symmetry
of the problem that, for the field $\opr$, \beq \langle \opa_1
\rangle = \sqrt{T} \mr \eeq (Remember that, e.g., $\opa_1$ is the
in-phase amplitude of the field in a specific temporal mode, while
$\opr_1$ is the in-phase amplitude operator of a continuous flux.)
Assume now that $\oprp$ has a much larger mean in-phase amplitude
than $\opr$. Filtering away the DC-component of the signal
(\ref{eq: hom signal}), and measuring the variance of the detector
signal after filtering the temporal mode, the detector signal
approximately becomes \beq \mrp \triangle \opr_1 \ast h(t) \eeq
Using the same procedure as we used to compute the mean photon
flux above, we see that the variance of this signal is $(\mrp)^2
T/4$. From this we can conclude that $\langle \triangle \opa_1^2
\rangle = 1/4$, since $T (\mrp)^2$ is the square of the mean local
oscillator in-phase amplitude (the amplitude we beat the
fluctuations against).

If we phase rotate $\opr$ with $\pi/2$ with the respect to
$\oprp$, and follow the same procedure to calculate $\langle
\opa_2 \rangle$ and $\langle \triangle \opa_2^2 \rangle$, we will
find that $\langle \opa_2 \rangle = 0$ and $\langle \triangle
\opa_1^2 \rangle = 1/4$. Hence, the results in the third column of
Table \ref{table: variances} are reproduced if Eq. (\ref{eq:
spectrum 1}) and (\ref{eq: spectrum 2}) are used.

\subsection{Expectation values of the laser output field}
\label{app: laser output}

If we now use (\ref{eq: laser spectrum 1}) and (\ref{eq: laser
spectrum 2}), modelling the noise properties of the output field
from a laser, we will get identical results as those in the
previous subsection for $\langle \hat{n} \rangle$, $\langle
(\triangle \hat{n})^2 \rangle$, $\langle \opa_1 \rangle$, $\langle
\opa_2 \rangle$, and $\langle \triangle \opa_1^2 \rangle$.
However, due to the spectrum (\ref{eq: laser spectrum 2}) of
$\opr_2$, the expectation value of $\langle \triangle \opa_2^2
\rangle =\infty$ independent of $T$.

To study this divergence, due to the diffusion of $\opr_2$, assume
that we homodyne the field $\opr$ in two orthogonal temporal modes
defined by two rectangle functions with durations from $t$ to $t +
T$, and $t+\tau$ to $t+\tau + T$, respectively, where $\tau \geq
T$. A schematic procedure to make such a measurement with the aid
of a flip mirror is depicted in Fig. \ref{Fig: Homodyne
measurement}. From (\ref{eq: linearized homodyne}) we find that
the measurement output will become $2 \mr [\triangle \opr_2(t) -
\triangle \opr_2(t + \tau)]\ast h(t)$. With the help of (\ref{eq:
laser spectrum 2}), the spectrum of the output fluctuations can be
computed to be \beq S_{Hom}(\Omega) = 4 \rm^2 \frac{\Omega^2 +
\gamma^2}{\Omega^2} \left [ \frac{ sin(\Omega \tau/2)  sin(\Omega
T/2)}{\Omega/2} \right ]^2 \label{Eq: Time delayed spectrum} \eeq
Transforming this function back to the time domain gives the
correlation function, and its value at the origin, the variance,
can be computed to be: \beq \mr^2 T \left[ 2 - T \tau
\gamma^2(1-\frac{T}{6 \tau}) \right ] . \label{Eq: Hom variance}
\eeq We see that since $\tau \geq T$, the variance will
approximately be $2 \mr^2 T$ (the shot noise associated with the
total detected photon number $2 \mr^2 T$) when $T \tau <
\gamma^{-2}$. That is, for measurements, even relative
measurements, within the laser coherence time $\gamma^{-1}$, the
laser output field in a succession of temporal modes will, each
one, approximately be in a coherent state if the first mode in the
succession is used as the reference. Note that this conclusion
holds irrespective of the numerical values of $T$, $\tau$, and
$\gamma^{-1}$ as long as $T <\tau < \gamma^{-1}$. This result
agrees with the analysis of van Enk and Fuchs \cite{Enk}.

\subsection{First order coherence of phase-locked lasers}
\label{appendix: first order }

The first order correlation function between two classical
electromagnetic fields is defined \beq \langle E({\bf \bar{r}}, t)
E'^\ast({\bf \bar{r}}', t')\rangle , \label{eq: first order func}
\eeq where ${\bf \bar{r}}$ and ${\bf \bar{r}}'$ denote two spatial
locations, and $t$ and $t'$ denote two times. In the following we
will restrict ourselves to single mode fields, and therefore, we
can suppress the spatial coordinates. Expressing the fields in
terms of amplitude and phase, and expanding the amplitude in its
(real and positive) mean $E_0$, and its fluctuation around this
mean $\Delta E(t)$, one gets \beq \langle E(t) E'^\ast(t')\rangle
\approx E_0 E_0' e^{i\omega_0 (t - t')} \langle
e^{i[\phi(t)-\phi'(t')]}\rangle , \label{eq: polar
decomposition}\eeq where we have dropped the term quadratic in the
the fluctuations $\Delta E$, and where we have assumed that the
amplitude- and the phase-fluctuations are uncorrelated. (The
latter assumption is not true in semiconductor lasers, where the
so-called $\alpha$-parameter characterizes the inverted-media
meditated correlations between amplitude and phase. For small
fluctuations this fact is inconsequential for what follows.) From
Eq. (\ref{eq: polar decomposition}) we see that as long as
$|\phi(t)-\phi'(t')| \ll 1$ rad, then the correlation function has
the modulus $|E_0 E_0'|$, indicating that the two fields are first
order coherent. Assume that the difference $|\phi(t)-\phi'(t')|$
is small. We can then take either field and make it our fiducial
reference, so that we work in a frame rotating with the angular
frequency $\omega_0$. We find that in this rotating frame, $E_0
\leftrightarrow r_0$, $\triangle E(t) \leftrightarrow \triangle
\rone$, and for small angles we have $\phi(t) \leftrightarrow
\triangle \rtwo/r_0$. Similar relations hold for the primed field.
{\it Hence, if we can show that for two locked lasers the relation
\beq |\triangle \rtwo/r_0-\triangle\rtwo'/r_0' |\ll 1 \label{eq:
condition}\eeq holds, then our assumptions above hold, and the
laser fields will be first order coherent.} For two lasers with
equal fields, $r_0 = r_0'$, the condition can be reformulated \beq
|\triangle\rtwo -\triangle\rtwo'| \ll r_0 . \label{eq: condition
2} \eeq Hence, the homodyne measurement beat signal is directly
relevant to the two fields' relative first order coherence.

From (\ref{eq: homodyne noise}) we can deduce that the noise
spectrum of $\triangle\rtwo -\triangle\rtwo'$ equals $4$, where we
have assumed a mirror reflectivity of 1/2 for the beam splitters
$BS2$ and $BS2'$. This corresponds to an equivalent photon flux
amplitude of the order 1 Hz$^{1/2}$, very much below a typical
laser. E.g., a single (transverse) mode laser emitting 1 mW
optical power at a wavelength of 500 nm has a photon flux
amplitude of $5 \cdot 10^7$ Hz$^{1/2}$ demonstrating that
(\ref{eq: condition 2}) is easily met. This proves the validity of
our approach.

\begin{figure}[htbp]
\infig{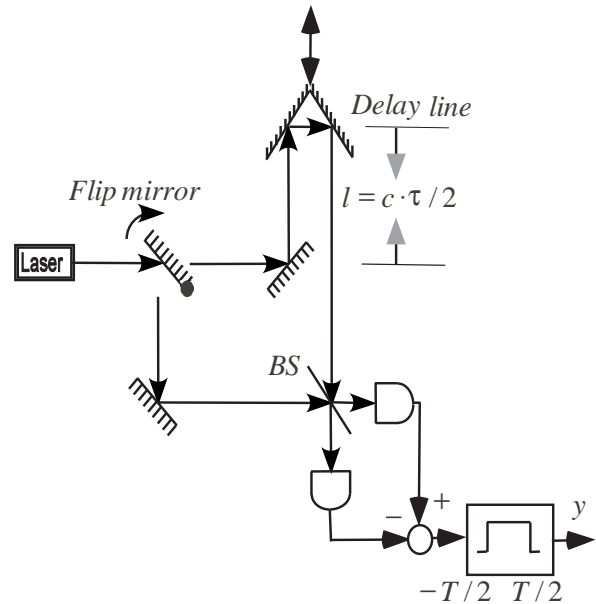}{8cm} \caption{Schematic drawing of a homodyne 
measurement of different temporal modes of a laser's output flux. 
The flip mirror is synchronized with the detector, and the 
separation between the temporal mode can be varied through the 
time delay $\tau=2 l /c$, where $l$ is the length of the 
time-delay ``trombone'' and $c$ is the phase-velocity.} 
\label{Fig: Homodyne measurement}
\end{figure}

\subsection{Laser potentials and the linearized approximation}

To see why the noise spectra (\ref{eq: laser spectrum 1}),
(\ref{eq: laser spectrum 2}) correctly predicts the short-time
($T$, $\tau < \gamma^{-1}$) relative-coherence properties of a
laser, but fails to reproduce the long term quadrature phase
properties (popularly speaking, these are called the coherence
properties) manifested in the first column in Table \ref{table:
variances}, it is instructive to look at the laser potential
models that underlies the various theories.
\begin{figure}[htbp]
\infig{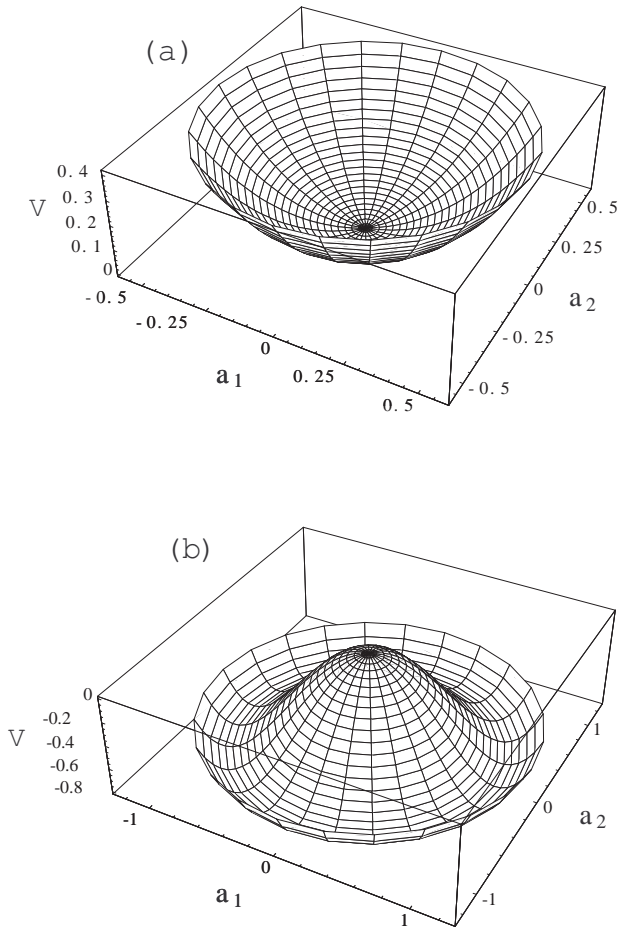}{\columnwidth}  \caption{A laser potential 
with $\gamma_0=1$ per unit time and $C=0.8$. (a) Below threshold 
pumping ($\alpha=0$). (b) Above threshold pumping 
($\alpha-\gamma_0=1.6$ per unit time).} \label{Fig: Laser 
potential}
\end{figure}

The internal field of a laser, under certain approximations can be
modelled as a particle in a particular potential subjected to
Langevin noise forces. The simplest standard model for the
potential, derived from the equation of motion of the field is
\beq V(\opa)=-(\frac{\alpha-\gamma_0}{4})|\opa|^2+
\frac{C}{4}|\opa|^4 , \label{eq: potential} \eeq where $\alpha$ is
the overall gain per unit time in the laser, $\gamma_0$ the cavity
decay rate (the inverse of the (cold) laser cavity photon
lifetime), and $C$ is the gain saturation parameter. In Fig.
\ref{Fig: Laser potential} we see how the potential goes from a
parabolic potential to a ``Mexican hat'' potential as the gain
goes from subthreshold to above threshold. In both cases the
potential has circular symmetry around the origin. Since lasing is
triggered by spontaneous emission (that is one of the origins of
the Langevin noise sources in the model), the laser field will not
have any preferred phase (relative to the fiducial reference
defining the coordinate system orientation). Therefore, a
free-running laser where only the intensity is known (supposedly
through a photon number measurement or by knowledge of the
relation between the pumping and the output intensity) must be
described by the density operator $ \oprho_L $ of Eq. (\ref{eq:
mixed state}). As M\o lmer \cite{Molmer}, and other's before him,
have pointed out, a laser does not induce any symmetry breaking,
and therefore it does not induce coherence relative to any other
oscillator.

However, as the analysis above indicate, by measuring and
influencing the field of a laser, either with feedforward control,
or feedback control, the relative phase of two lasers can be
locked so that the lasers become relatively coherent. However, if
the locking servo is turned off, the Langevin fluctuations of each
laser will make the relative phase between the two lasers diffuse,
since the potential offers only a restoring force in the radial
direction, but not in the azimuthal direction. After a time
approximately equal to the inverse of the laser emission spectral
linewidth, the two lasers are no longer relatively coherent.

This situation is not well described by Eqs. (\ref{eq: laser
spectrum 1}) and
 (\ref{eq: laser spectrum 2}), because these equations predict that
it is only the $\atwo$ quadrature that diffuses. If that were the
case, then the relative phase of the two lasers with the initial
relative phase zero, would remain zero, on average, for all
subsequent times. If so, the quadrature-phase amplitude
fluctuation would eventually become larger than the mean in-phase
component. However, this is not what happens in a laser. The
reason for the erroneous prediction of (\ref{eq: laser spectrum
2}) is that in the analysis leading to the equation, the laser
potential (or rather the Langevin noise operators) have been
linearized about the operating point. The corresponding linearized
potential is illustrated in Fig. \ref{Fig: Linearized laser
potential}. One can see that in the linearized potential, the gain
saturation will only act in the direction parallel to the field
expectation value. In contrast, the better model of the potential
will act in the direction parallel to the instantaneous field,
i.e., in the radial direction. The linearized potential correctly
predicts the laser behavior for times up to the laser's relative
decoherence time, but fails for long time scales, where it is
actually the relative phase, and not the quadrature field
amplitude that diffuses.
\begin{figure}[htbp]
\infig{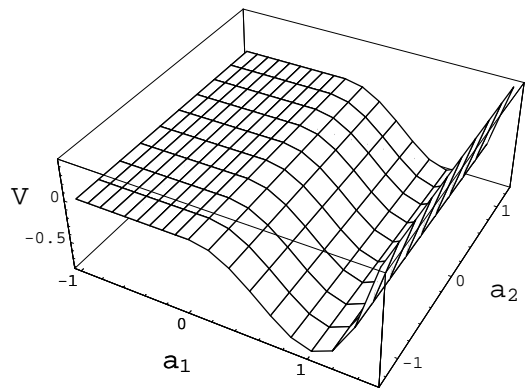}{7cm} \caption{A linearized laser potential 
above threshold} \label{Fig: Linearized laser potential}
\end{figure}


\begin{table}
\begin{ruledtabular}
\begin{tabular*}{\hsize}{l@{\extracolsep{0ptplus1fil}}c@{\extracolsep{0ptplus1fil}}c@{\extracolsep{0ptplus1fil}}c@{\extracolsep{0ptplus1fil}}}
 &$\hat{\rho}_L$&$\ket{e^{i \phi}\alpha}$&Laser model\\
\colrule
$\langle \aone \rangle$&0&$|\alpha| $&$\sqrt{T} \mr$\\
$\langle \atwo \rangle$&0&$0$&$0$\\
$\langle \Delta \aone^2 \rangle$&$\frac{2|\alpha|^2 + 1}{4}$&1/4&$1/4$\\
$\langle \Delta \atwo^2 \rangle$&$\frac{2|\alpha|^2 + 1}{4}$&1/4&$\infty$\\
$\langle \hat{n} \rangle  = \langle \hat{a}^\dagger \hat{a} \rangle$&$|\alpha|^2$&$|\alpha|^2$&$T \mr^2$\\
$\langle (\triangle \hat{n})^2 \rangle $&$|\alpha|^2$&$|\alpha|^2$&$T \mr^2$\\
\end{tabular*}
\end{ruledtabular}
\caption{A table comparing the predicted statistics of three
different fields. A totally statistically mixed state with
Poissonian statistics, a coherent state, and the predicted output
of a laser using Langevin equations.} \label{table: variances}
\end{table}

\end{document}